\documentclass[epj]{svjour}
\usepackage{graphicx}
\usepackage{bm}
\usepackage{epsfig}

\def\OMIT#1{}
\newcommand{\ltap}{\stackrel{<}{_\sim}}
\newcommand{\gtap}{\stackrel{>}{_\sim}}

\newcommand{\bea}{\begin{eqnarray}}
\newcommand{\eea}{\end{eqnarray}}

\begin{document}
\title{\boldmath Power corrections in $e^+ e^-  \to {\pi^+  \pi^-,\,K^+ K^- } $ and $B\to K \pi,\,\pi \pi$}
\author{Murugesh Duraisamy\inst{1}
\thanks{\emph{Present address:} Department of Physics and Astronomy, University of Mississippi, University, MS, 38677} \and Alexander L.~Kagan \inst{1}
}                     
%
%
\institute{Department of Physics, University of Cincinnati, Cincinnati, OH 45221}
\date{Received: date / Revised version: date}
%
\abstract{
CLEO-c measurements of the timelike form factors $F_{\pi}$, $F_{K}$
at $\sqrt{s}=3.671$ GeV provide a {\it direct} probe of power corrections (PC's) at energies near $m_B$.  
PC's in $F_{\pi,K}$ and $B \to K\pi\,,\pi\pi $ are separated into perturbative 
and soft parts. In $F_{\pi,K}$ the latter
are $\ge O(10)$ larger. A PC fit to the
$B\to K\pi,\pi\pi$ data also yields a $\ge O(10)$ soft-to-perturbative hierarchy for
the QCD penguin PC's. Hence, both can be attributed to dominance of the soft-ovelap between energetic (approximately) back-to-back collinear partons, and
consistency of the $B\to K\pi,\pi\pi$ fit with the Standard Model appears to be naturally realized.  
The CP asymmetries $S_{K_s \pi^0}$, $C_{K_s \pi^0}$ are well determined, providing a clean test for new physics.    
\PACS{
      {
      13.25.-k
      }{
      hadronic decays: mesons
      }
       \and
     {
     13.66.Bc
     }{
     hadrons: production by electron-positron collisions
     }
     } 
} 
\maketitle
Much effort has gone into the theoretical description of $B$ decays into light meson pairs.
Apart from being of interest in QCD, the issue has important implications for new physics search strategies which rely on 
comparing decay rates and CP asymmetries in different final states.
The decay amplitudes can be organized into expansions in powers of $1/m_b$. 
The leading power (LP) contributions are calculable in QCD factorization (QCDF)~\cite{BBNS1} in terms of universal non-perturbative quantities.  
Numerous leading power predictions for $B \to M_1 M _2 $  decays are in gross conflict with the data.  
In $B \to K \pi,\,\pi \pi$ the direct CP asymmetry $A_{\pi^+ \pi^- }$ is too small, $A_{K^+ \pi^- }$ is too small and of wrong sign, $A_{K^+ \pi^- } \approx A_{K^+ \pi^0}$, contrary to observation, and the branching ratios ${\rm Br}_{K^0 \pi^0 }$,  ${\rm Br}_{\pi^0 \pi^0 }$ are too small.  A possible explanation is that certain power corrections (PC's) are of same order as
or larger than their LP counterparts and have large strong phases, due to non-perturbative effects.

Continuum $e^+ e^- \to M_1 M_2 $ light meson cross sections
at $\sqrt{s} \approx 3.7$ and $10.58$ GeV at the charm and $B$ factories
provide a direct probe of PC's in the timelike vector-current matrix elements, $\langle M_1 M_2 | \bar q \gamma_\mu q |0 \rangle$.  
Perturbative calculations of 
PC's on the light-cone
contain IR log-divergent terms of the form
$ \alpha_s  (\mu_h) (1/\sqrt{s})^n \ln^m (\sqrt{s} /\Lambda )$, signaling the breakdown of short/long-distance factorization (substitute $\sqrt{s} \to m_B$ in $B$ decays).  
$\Lambda$ represents a physical IR cutoff on the longitudinal momentum of, e.g., a valence quark, in the convolution integrals of light meson light-cone distribution amplitudes (LCDA's) with hard scattering amplitudes.
We therefore divide the PC's into perturbative and non-perturbative parts (soft overlaps), where  the former are defined by imposing 
$\Lambda, \mu_h \gtap 1 $ GeV.
For example, the vector-current form factor PC's are written as
$\delta F\, = \,\delta F^{\rm pert.} + \delta F^{\rm n.p.}$.
The leading kinematic final-state parton configurations responsible for the non-perturbative (or perturbative) parts of the vector-current and penguin PC's are similar.  Thus, we may learn about the relative importance of soft-overlaps, e.g. end-point effects, in the latter from the continuum data. 



The continuum timelike form factors $F_{K}\,,F_{\pi}$ measured by CLEO-c at $\sqrt{s}= 3.671$ GeV are \cite{CLEOc}
\begin{equation}
|F_\pi |= 0.075 \pm 0.009\,,~~~|F_K |= 0.063 \pm 0.004\,.
\label{FPexp}
\end{equation}
The calculable LP contributions arise at twist-2 in the LCDA's, and fall like $1/s$~\cite{brodskylepage}.
We obtain 
\begin{equation}
F_\pi^{\rm LP}= {-0.01}^{+0.002}_{-0.004 }\,,~~F_K^{\rm LP}={-0.014}^{+0.002}_{-0.006}
\label{twist2numerics}
\end{equation}
at tree-level.  The errors are due to variation of the first two LCDA Gegenbauer coefficients~\cite{ballLCDA} and the scale 
$\mu\in [\sqrt{s}/2 ,\sqrt{s}]$
at which they and $\alpha_s$ are evaluated, added in quadrature.
The `central-values' are for $\mu= \sqrt{s}$, and $\alpha_s$ is evaluated at two-loops throughout this work.
Even at $\mu= 1$ GeV, $F^{\rm LP}_\pi \,(F^{\rm LP}_K) \approx -0.025\,(-0.036)$, 
implying that
$F_K$ and especially $F_\pi$ 
are dominated by PC's.



\begin{figure}[tl]
\centerline{
\hspace{-0.5cm}\vbox{
\hspace{-.054cm}\includegraphics[width=7.564truecm,height=3.5truecm]{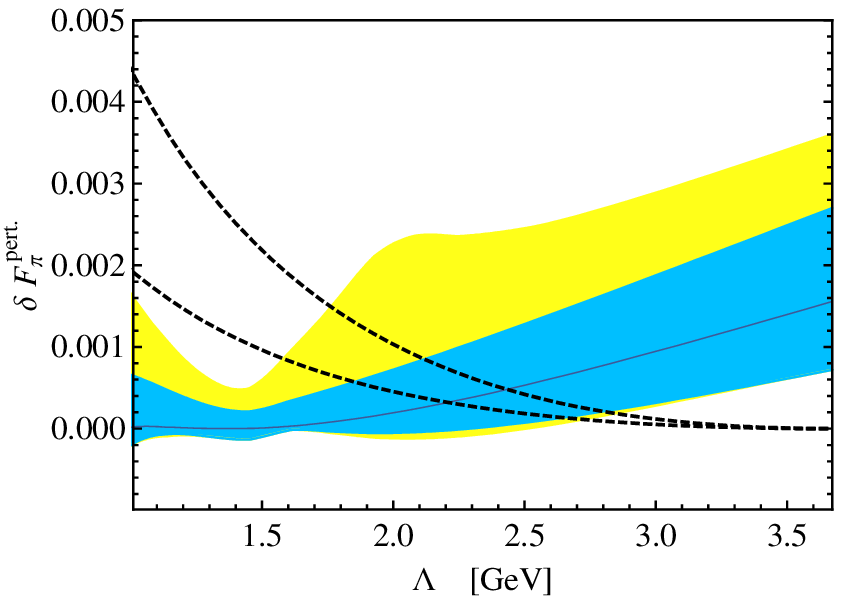}\vspace{-.565cm} 
\hbox{\hspace{-.06cm}\includegraphics[width=7.57truecm,height=3.5truecm]{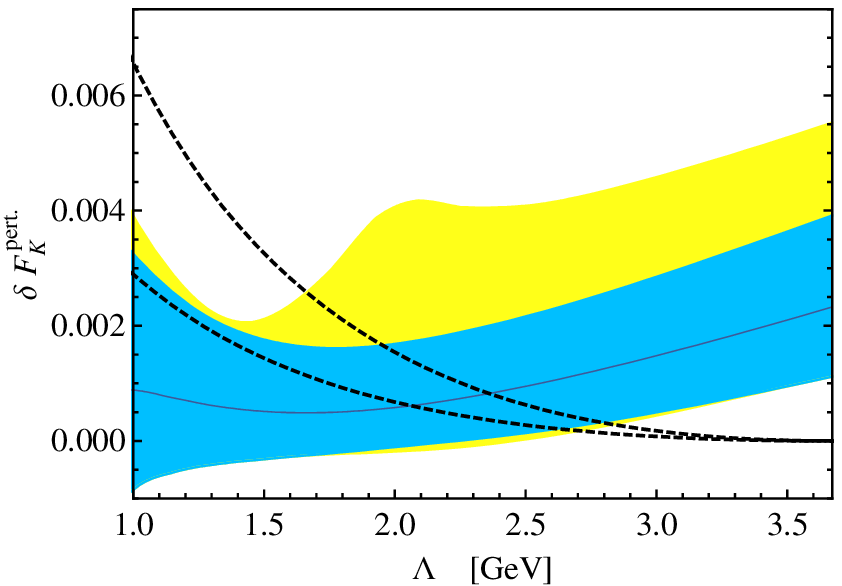}}} }
{\caption[1]{\label{fig:FPvsLambdacut}
$\delta F^{\rm pert.}_{\pi ,K}$ vs. $\Lambda$:
solid curves for $\alpha_s $, central values of LCDA parameters evaluated at 
$\mu_h = \Lambda$;
inner blue bands for LCDA parameters varied within errors;
outer yellow bands include $\mu_h \in [{\rm Max}[1\,{\rm GeV},\, \Lambda/2 ]\,, \sqrt{s}]$; dashed lines outline `outer bands' for asymptotic LCDA's.
Errors added in quadrature. }} 
\end{figure}



$\delta F_{K,\pi}$ enter at $1/E^2 $, or twist-4 perturbatively, 
and to first approximation fall like $1/s^2$.
We obtain $\delta F_{K,\pi}^{\rm pert.}$ from convolutions of two twist-3 valence quark LCDA's with the 
tree-level hard-scattering amplitudes (twist-4 valence quark LCDA's contribute negligibly).   
The model parameters of~\cite{ballLCDA} are employed for the LCDA's.
Perturbative higher Fock state effects are of same twist and order of magnitude, and therefore would not qualitatively alter our conclusions.
Fig.~1 shows the ranges obtained as the cutoff $\Lambda$ in the divergent terms is varied
from $\sqrt{s}$ to 1 GeV.  $\Lambda$ is roughly the lowest gluon virtuality allowed.  
Results for asymptotic LCDA's are also shown.
There are large accidental cancelations between asymptotic and non-asymptotic effects at lower $\Lambda $. The magnitudes of each separately therefore give a better indication of the size of perturbative effects for $\Lambda$ near 1 GeV.
Comparing the asymptotic plots to Eq.~\ref{FPexp},  it is clear that 
the dominance of the PC's in $F_{K,\pi}$ is due to their soft parts,
\begin{equation}
|{\delta F_\pi^{\rm n.p.} /  \delta F_\pi^{\rm pert.}}| \ge O(10),~~~|{\delta F_K^{\rm n.p.} /  \delta F_K^{\rm pert.}}| \ge O(10).
\label{eq:nonpertvspert}
\end{equation}
Similar soft enhancement would account for $F_\pi (m_{J/\Psi} ) \approx 0.10$, as extracted from $J/\Psi$ decays \cite{milana}.  
At LP, the form factors obey canonical $SU(3)_F$ flavor symmetry breaking, i.e.,  $(F_\pi / F_K )_{\rm LP}  \approx f_\pi^2 /f_K^2 \approx 0.7 $.
However, $|F_\pi /F_K |_{\rm exp.}=1.20 \pm 0.17$, implying that
$|\delta F_\pi /\delta F_K  |> 1$, e.g., $(1.9\pm 0.36)\, f_\pi^2 /f_K^2$ for constructive interference
between LP and PC effects.
Apparently,
the soft-overlap is significantly larger for $\pi\pi$ than $KK$.



The $SU(3)_F $ diagrammatic representation gives a convenient general classification
of the  $B \to K\pi,\, \pi\pi$ amplitudes \cite{gronaurosner1}.   
For example, the $\bar B \to K \pi$ amplitudes are
\begin{eqnarray}
A_{\bar{K}^0 \pi^- } &=& \lambda_p  (A \,\delta_{p\,u}  + P^p -{1\over 3} P_{\rm EW}^{{\rm C}\,,p  }+{2\over3} P_{\rm EW}^{{\rm E}\,,p}  )\nonumber\\
-A_{K^- \pi^+  } &=& \lambda_p ( T \,\delta_{p\,u}  + P^{p}   +{2\over 3} P_{\rm EW}^{{\rm C}\,,p  } -{1\over3} P_{\rm EW}^{{\rm E}\,,p} )\nonumber\\
-\sqrt{2} A_{K^- \pi^0 } &=&\lambda_p ( (T+ C+ A )\,\delta_{p\,u} +  P^p+P^p_{\rm EW}\nonumber \\
 & + &{2\over3} P_{\rm EW}^{{\rm C}\,,p  }+{2\over3} P_{\rm EW}^{{\rm E}\,,p}  ) \nonumber\\
 \sqrt{2}  A_{\bar{K}^0 \pi^0} &=& A_{\bar{K}^0 \pi^- }+\sqrt{2} A_{K^- \pi^0 } - A_{K^- \pi^+ }  \,.
 \label{eq:Kpiamps}
\end{eqnarray}
The CKM factor $\lambda_p\! =\! V_{p b} V^*_{p s} $, and there is a sum over $p=u,c$.  The $B^-\,\,(B^0)$ Br's are given by
$|A|^2\,\,(\tau_{B^0 } / \tau_{B^-} |A|^2 )$. 
$T \,(a_1 )$ and $C \,(a_2)$ are the color-allowed and color-suppressed `tree' amplitudes, and $P^{p} \,(a_{4,6} ), P_{\rm EW}^{p} \,(a_{7,9})$, and $P_{\rm EW}^{{\rm C}\,,p} \,(a_{8,10} )$ are the QCD penguin, electroweak penguin (EWP), and color-suppressed EWP amplitudes, respectively.  
They consist of LP parts $T_{\rm LP}$, {\it etc.} (the QCDF coefficients $a_i $~\cite{BBNS1}
are in parenthesis) and PC's $\delta T $, {\it etc.}
The corresponding $\pi\pi$ amplitudes are primed.
$P_{\rm EW}^{{\rm E}\,,p}$ ($P_{\rm EW}^{\prime\,{\rm E}\,,p},\,P_{\rm EW}^{\prime\,{\rm A}\,,p}$)  
and $A$ ($E^\prime$) are the $K\pi$ ($\pi\pi$) EWP and `tree' weak annihilation PC's, respectively.
We can neglect the electromagnetic $u,c$-loop penguin contractions in the EWP PC's, and thus drop
their `$p$' superscripts below.

\begin{figure}[tr]
\centerline{
\hspace{-0.4cm}\vbox{
\hbox{\includegraphics[width=4.0truecm,height=2.0truecm]{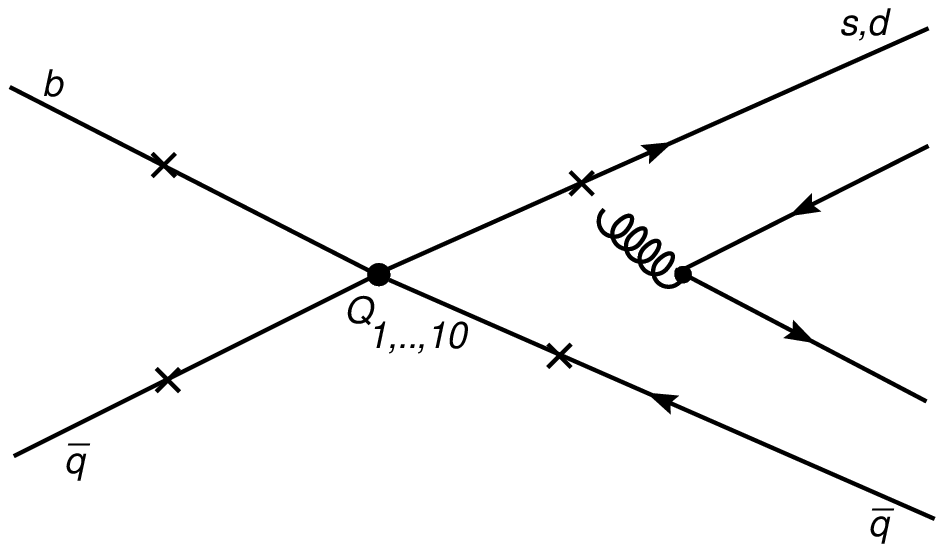}\hspace{0.2cm}\includegraphics[width=4.0truecm,height=2.0truecm]{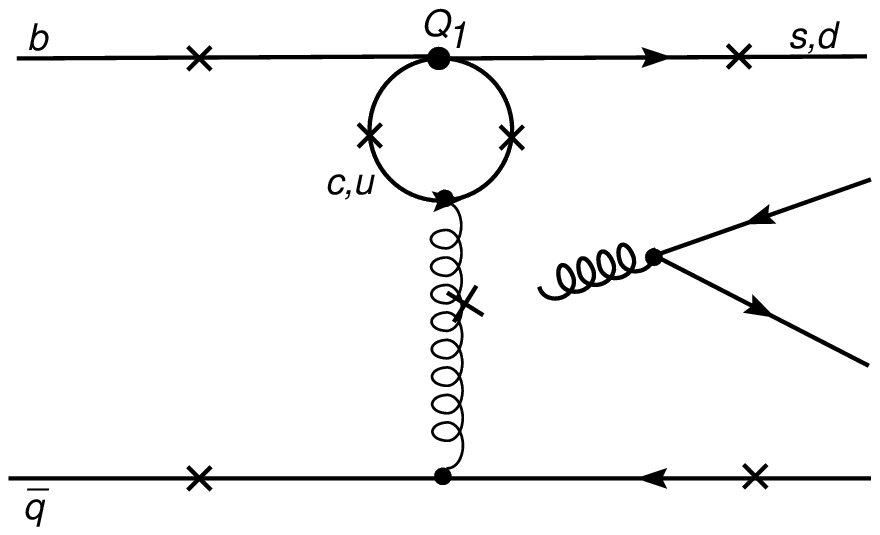}}\vspace{0.2cm} 
\hbox{\includegraphics[width=4.0truecm,height=2.0truecm]{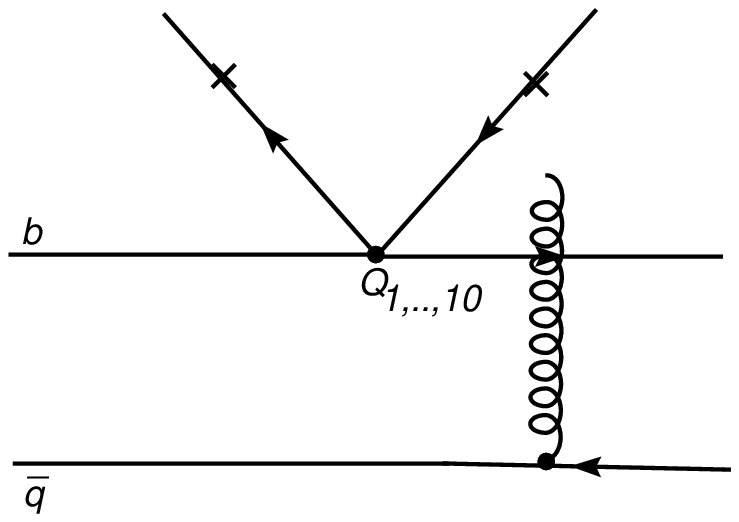}\hspace{0.2cm} 
\includegraphics[width=4.0truecm,height=2.0truecm]{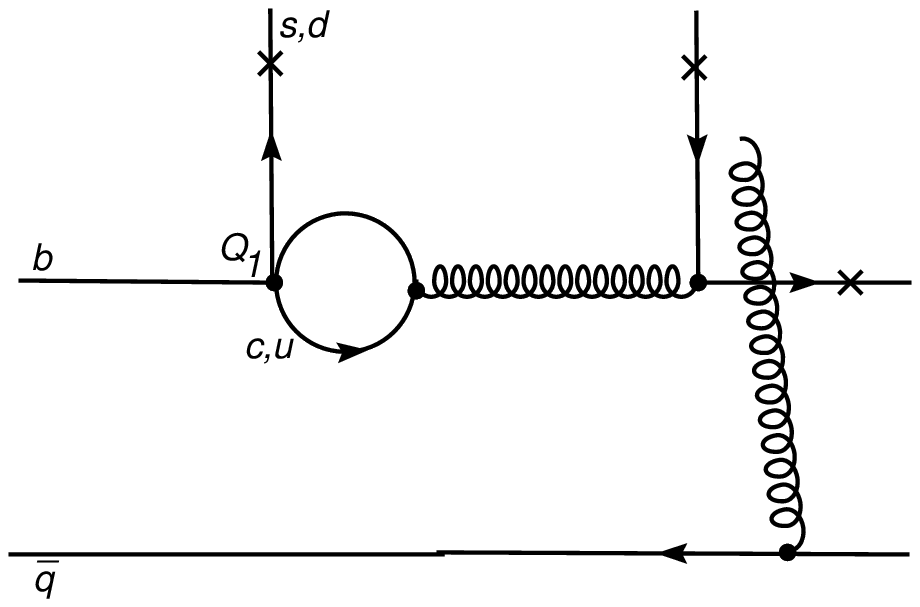}}\vspace{0.2cm} 
\hbox{\includegraphics[width=4.0truecm,height=2.0truecm]{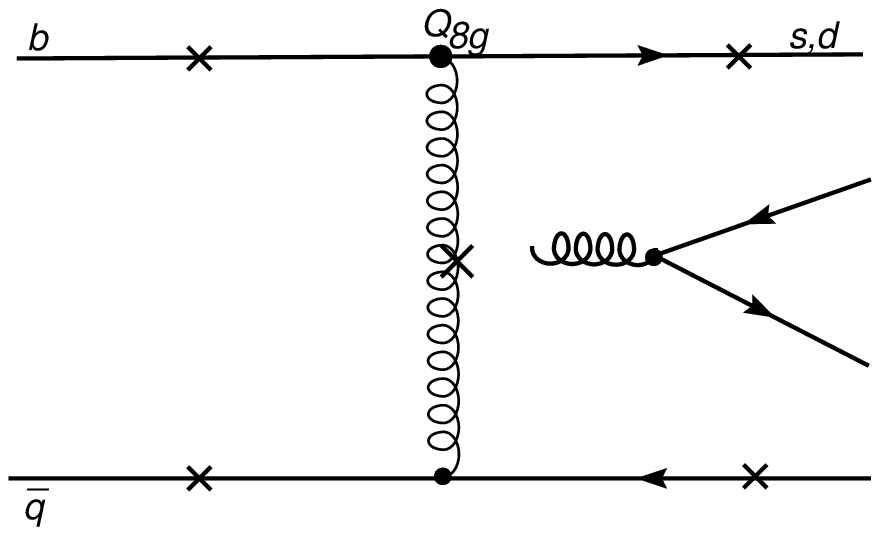}\hspace{0.2cm}\includegraphics[width=4.0truecm,height=2.0truecm]{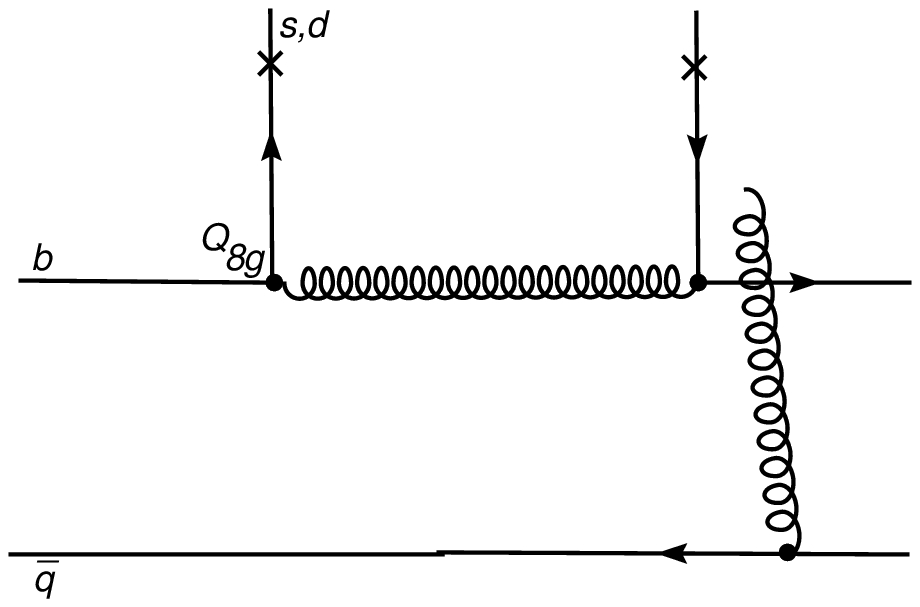} }}}
{\caption[1]{\label{fig:FeynDiags}
Diagrams for perturbative $B\to M_1 M_2$ PC's. Crosses indicate all places where the gluon can attach.  The  
$Q_i$ are the $\Delta B=1$ effective Hamiltonian operators \cite{BBNS1}.}} 
\end{figure}


\begin{figure}[tr]
\centerline{
\hspace{-0.5cm}\vbox{
\hspace{-.15cm}\hbox{\hspace{0.2cm}\includegraphics[width=7.220truecm,height=3.55truecm]{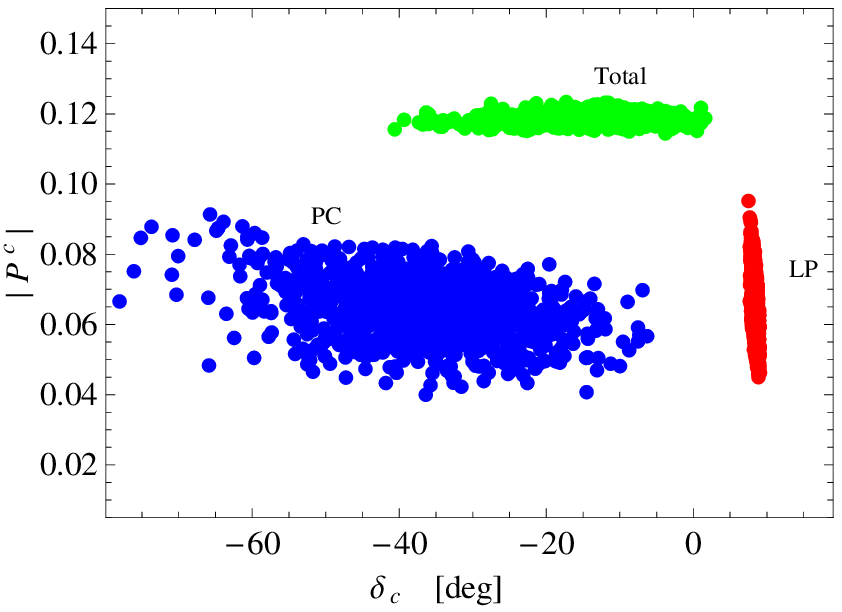}}\vspace{-.605cm}
\hbox{\includegraphics[width=7.254truecm,height=3.55truecm]{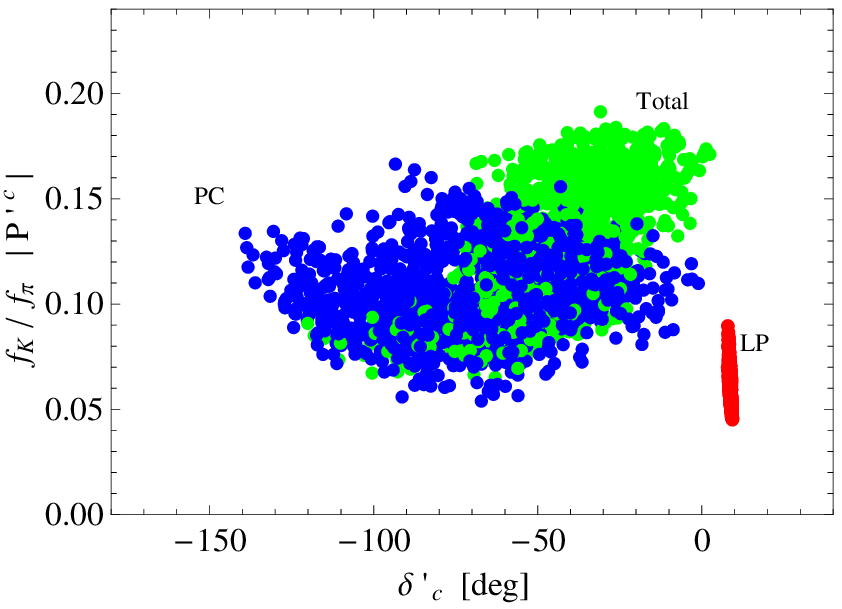}}\vspace{-.595cm} 
\hbox{\hspace{0.19cm}\includegraphics[width=7.07truecm,height=3.55truecm]{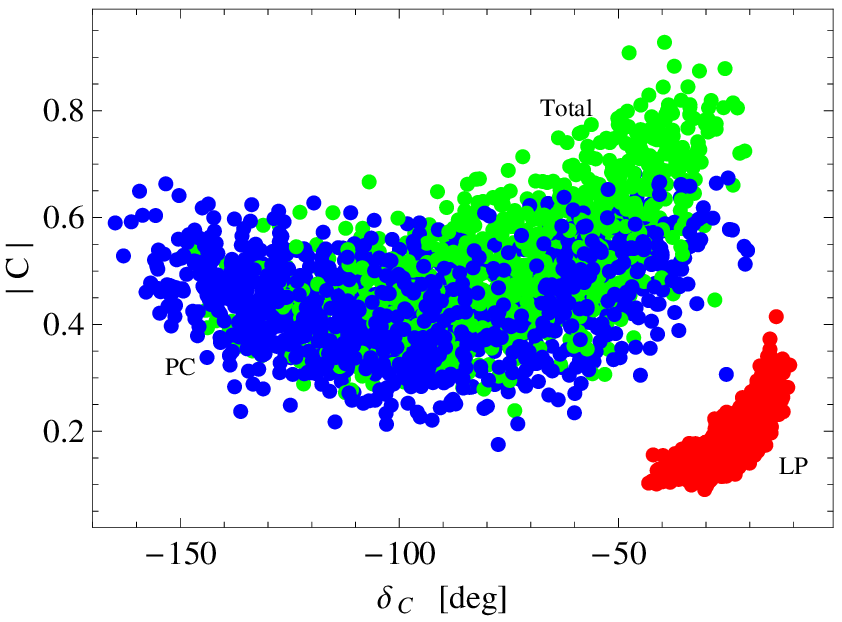}}\vspace{-.58cm} 
\hspace{0.2cm}\hbox{\hspace{0.10cm}\includegraphics[width=7.37truecm,height=3.55truecm]{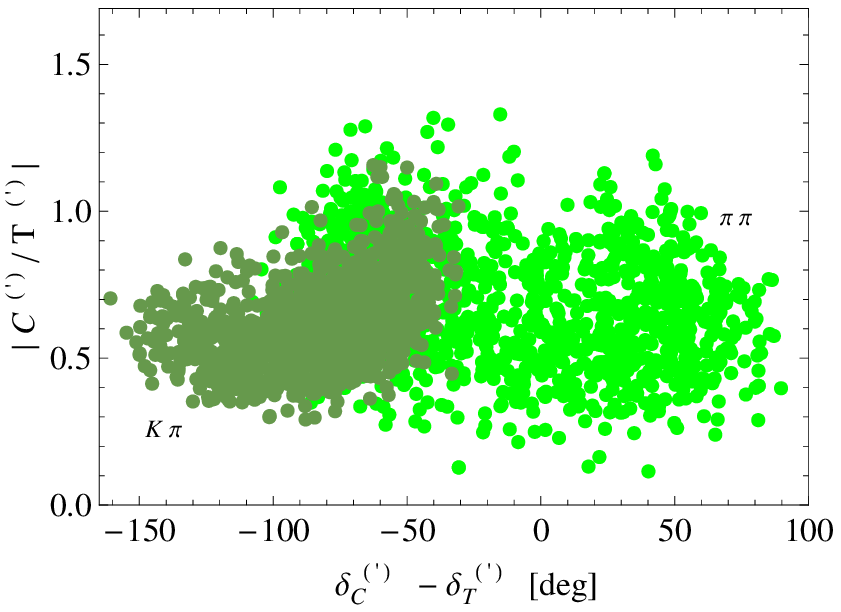}}\vspace{-.68cm}
\hbox{\hspace{0.005cm}\includegraphics[width=7.39truecm,height=3.55truecm]{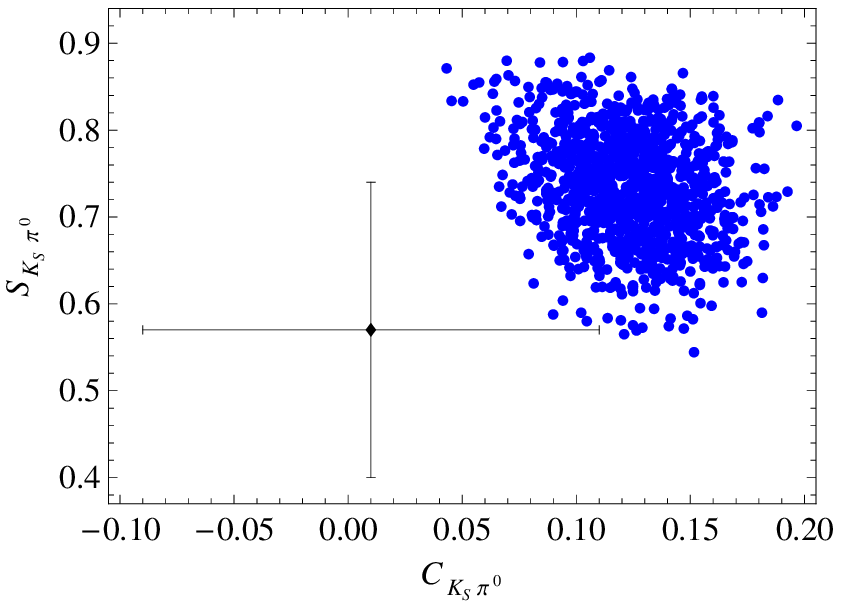}}}}
{\caption[1]{\label{fig:KPiPCfits}
$K\pi$, $\pi\pi$ fits: (a) $\delta P^c $ (blue), $P^c_{\rm LP}$ (red), $P^c $ (green), strong phases relative to naive factorization;
(b) same for $\delta P^{\prime\,c} $, 
$P^{\prime \,c}_{\rm LP}$, $P^{\prime\,c} $;
(c) same for $\delta C $, $C_{\rm LP} $, $C $;
(d) $|C/T |$ (olive green), $|C^\prime /T^\prime |$ (green) vs. strong phase differences;
(e) $S_{K_s \pi^0} $, $C_{K_s \pi^0 }$ for $\beta\in  [20.2,22.0^\circ ]$, and HFAG averages.}}
\end{figure}

The ${\rm Br}$'s and 
$A_{K^+ \pi^-}$, $A_{{K}^0 \pi^+}$, $A_{K^+  \pi^0}$, $A_{\pi^+\pi^-}$, $S_{\pi^+ \pi^- }$ 
are required to lie within their 1$\sigma$ HFAG experimental errors \cite{KPiHFAG}.
The LP amplitudes are evaluated in QCDF to NLO \cite{BBNS1}.  
The NNLO corrections \cite{benekejager1,benekejager2} would not have a substantial impact on our fit results. 
The LP inputs are varied uniformly 
within their errors.  The Wilson coefficients, $\alpha_s $, and the LCDA parameters \cite{ballLCDA,BBNS1}
are evaluated at the scale $\mu_b  \in [m_b /2\,,m_B]$, with $m_b=4.2$ GeV,
$m_c=1.3$ GeV,  $m_s =  100 \pm 20$ MeV, $f_B= 220\pm 20$ MeV, 
$F_{ B\to \pi} = 0.23 \pm 0.04$, $\lambda_B= .35\pm.15$ GeV, $V_{ub}= (3.86\pm .10) \cdot 10^{-3}$, $V_{cb}=0.041$, and $\gamma \in [50^\circ \,,\,80^\circ ]$.
The $K\pi$ ($\pi\pi$) PC fits are
dominated by $|\delta P^{(\prime) \,p} |e^{i \delta^{(\prime)}_p}$, $|\delta C^{(\prime)} | e^{i \delta^{(\prime)}_C }$, $|\delta T^{(\prime)} | e^{i \delta^{(\prime)}_T }$,
with strong phases defined relative to the corresponding 
naive-factorization amplitudes.
The fits allow $\delta P^{(\prime)\,u }$ and $\delta P^{(\prime)\,c } $ to differ substantially, a possibility suggested by their perturbative contributions, see Fig.~4c.
We require
$ |\delta P^{(\prime)\,u} /\delta P^{(\prime)\,c} | \le 3$; $|\delta T/\delta C |\le 0.4$ (we allow for $O(1)$ variation of the one-gluon exchange approximation
ratio, $\approx |C_2 /C_1 |$);
and $|A|,|E^\prime|< O(10) \,\times\,$their perturbative ranges, see below.
To good approximation, the EWP PC's satisfy 
\begin{eqnarray}
\label{eq:ewksu3relns}
{4\over 3} \delta P_{\rm EW}^{(C)}=\sum_{\kappa =\pm 1}\left({C_{10 (9)}+\kappa \,C_{9 (10)}  \over C_1 +\kappa \,C_2 } \right)(\delta T+\kappa\,  \delta C) 
\end{eqnarray}
in the $SU(3)_F$ limit \cite{Gronau:1998fn} (and similarly for $\pi\pi$),
and the order of magnitude one-gluon exchange approximation relations
$P^{\rm E}_{\rm EW } \sim 3/2\, C_9 /C_2 \, A $, $P_{\rm EW}^{\prime\,{\rm E}} \sim 3/2\, C_9 /C_1 \, E^\prime $, and 
$P_{\rm EW}^{\prime\,{\rm A}} \sim 3/2\, C_{10} /C_1 \, E^\prime $ \cite{BBNS1}, where
the $C_i$ are Wilson coefficients in the $\Delta B=1$ effective Hamiltonian.
The last three relations imply that $P^{(\prime)\,{\rm E}}_{\rm EW }$ and $P_{\rm EW}^{\prime\,{\rm A}}$ are negligible.
The $SU(3)_F$ breaking corrections in Eq. (\ref{eq:ewksu3relns}) could be $O(1)$ for $K\pi$, thus
multiplicative factors $r_{(C)} e^{i \delta_{(C)}}$ are introduced on the r.h.s., with $r_{(C)}  \in[0,2]$, $\delta_{(C)} \in [0,2 \pi]$.  


The fits yield 
$|\delta P^c | \sim |P^{c}_{\rm LP}|$, $|\delta P^{\prime\,c}| \gtap |P^{\prime\,c}_{\rm LP}|$, 
$|\delta C^{(\prime)} |\gtap | C^{(\prime)}_{\rm LP} |$ and $\delta_C \gtap 30^{\circ}$, see Figs.~3a-c.
A breakdown of the $1/E$ expansion is not implied if
a PC exceeds its LP counterpart (it could be accidental given that one is factorizable and the other is not),
but rather if power counting for the PC's themselves is violated, for which there is 
no indication. 
In fact, the continuum $e^+e^- \to \rho \eta$ cross sections at $\sqrt{s} \approx 3.77$ and 10.58 GeV give remarkably precise confirmation of the power counting rules \cite{inprep}.
In Fig.~3b, $P^{\prime \,c}$ is multiplied by 
$f_K / f_\pi $ for comparison with Fig.~3a.
Canonical $SU(3)_F$ breaking at LP gives $f_K /f_\pi  \,P_{\rm LP}^{\prime\,c}  \approx P_{\rm LP}^c$.
However,  it appears that $f_K /f_\pi  \,\delta P^{\prime\,c} > \delta P^c$, in accord with 
$\delta F_\pi  > \delta F_K$.
The magnitudes of $C^{(\prime)}/T^{(\prime)} $ in Fig.~3d can be smaller 
than in $SU(3)_F$ fits, see e.g. \cite{chiang}.
The need for a large strong phase difference $\delta_C -\delta_T$ is well known.
In terms of their experimental errors, $\delta P_{\rm EW}^{{\rm C}} $ can shift ${\rm Br}_{K^+ \pi^- }$, ${\rm Br}_{K^+ \pi^0 }$, and $A_{K^+ \pi^- }$ by $\ltap 2.5\sigma,1.5\sigma$, and $1.5 \sigma$, and $E^\prime$ can shift 
${\rm Br}_{\pi^+ \pi^- }$, $A_{\pi^+ \pi^-}$, and $S_{\pi^+\pi^-}$ by $\ltap 5\sigma$ ($20\%$), $2\sigma$, and $1.5\sigma$, respectively.
Other shifts due to subleading amplitudes are $< 1\sigma$.
The SM predictions for the time-dependent $CP$ asymmetries $S_{K_s \pi^0}$ and  $C_{K_s \pi^0 }= -A_{K_s \pi^0}$  in Fig.~3e
are consistent with experiment.
The experimental errors exceed the fit errors,
making this a good place to look for new physics \cite{fleischerzupan,gronaurosner2}.
The predicted  ranges for $A_{\pi^+ \pi^0}$ and $A_{\pi^0 \pi^0}$ are $\approx  [-0.06,+0.06]$ 
and $[-0.95,0.55]$, respectively, 
consistent with the HFAG averages, $0.06\pm 0.05$ and $0.43\pm0.25$.

To obtain an approximate goodness of fit for $K\pi$, we find the minimum for the scatter points of a $\chi^2$ constructed from
the ${\rm Br}$'s, direct CP asymmetries,  $S_{K_s \pi^0}$, $\beta = (21.1\pm 0.9)^\circ$ \cite{KPiHFAG}, and $\gamma = (67.8\pm 4.2)^\circ$ \cite{CKMFitter}.
The effective number of parameters fit is 9: $|\delta P^{c,u}|$, $|\delta T  |$, $|\delta C  |$, 3 strong phases, $\gamma$ and $\beta$  (the dependence of $\chi^2{\rm min} $ on $r_{(C)}$, $A$, and $P_{EW}^E$ is negligible).  The result, $\chi^2_{\rm min}/{\rm d.o.f}  \approx 3.5/2$, is consistent with 
the SM at $\approx 1.4 \sigma $ (83\% CL).  
To check that the pattern of PC's is natural, we compare the
soft-to-perturbative PC ratios to those in Eq. (\ref{eq:nonpertvspert}).


The perturbative $B \to K \pi, \,\pi\pi $ PC's are obtained from the diagrams of Fig.~2.
They depend on two renormalization scales:
(i) $\mu_b $, linked to the energy release of the decay, as in the LP amplitudes, and (ii) $\mu_h$, linked to the IR cutoff $\Lambda$.
The Wilson coefficients and one $ \alpha_s  $ factor in Figs.~2b,d 
(associated with the $u,c\,$-loops)
are evaluated at $\mu_b $.  The other $ \alpha_s $ factors and the LCDA parameters are evaluated at $\mu_h$.
The Wilson coefficients in Figs.~2a,c,e are NLO, and $C_1$ is LO in Figs.~2b,d.
We have checked that the quark loop diagrams in Fig.~2 eliminate the dominant leading $\log \mu_b$ scale dependence ($\propto C_1 \alpha_s /\pi$) in $\delta P^{p,\,{\rm pert.}}$.
Products of twist-2,3$\,\times\,$twist-2,3 
$K$,$\pi$ valence quark LCDA's are included in the 
amplitudes.
Our results are summarized 
in Fig.~4.  The largest contributions to $\delta P^{(\prime)\,p\,,{\rm pert.}} $ come from 
the charm-loop diagrams in Fig.~2b, the dipole operator ($Q_{8g}$) in Fig.~2e, and QCD penguin operator ($Q_{3,..,6}$) weak annihilation in Fig.~2a. 
They are dominated by contributions  
in which a gluon does not attach to the $B$.



\begin{figure}[tl]
\centerline{
\hspace{-0.5cm}\vbox{
\hspace{-.066cm}\includegraphics[width=7.564truecm,height=3.5truecm]{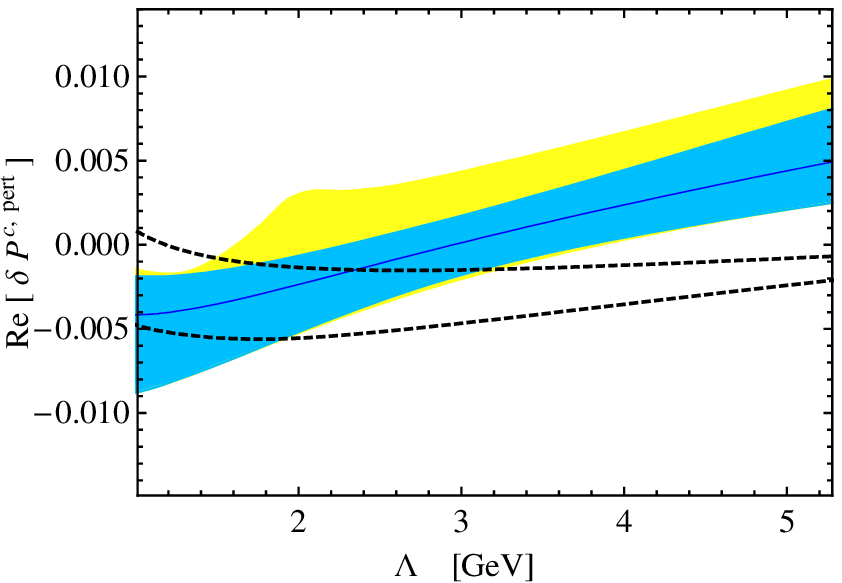}\vspace{-.57cm} 
\hbox{\hspace{-.075cm}\includegraphics[width=7.564truecm,height=3.5truecm]{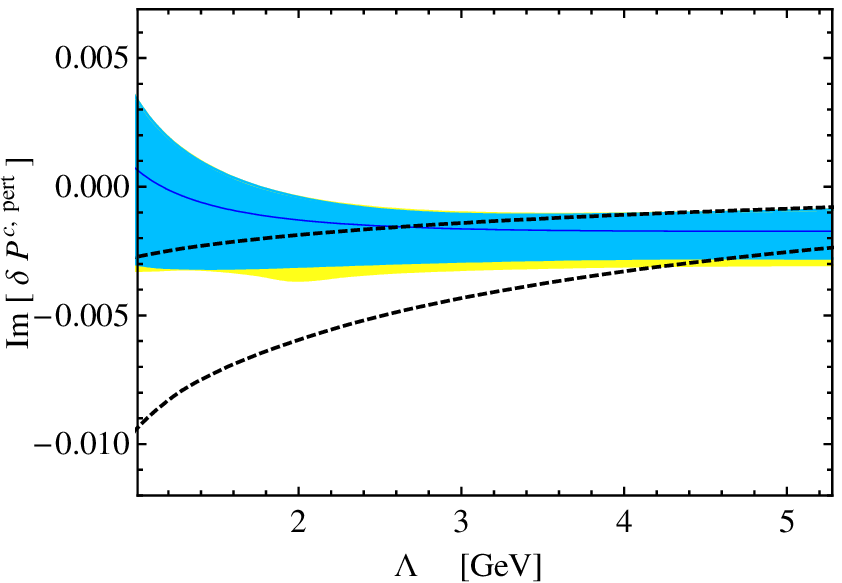}}\vspace{-.585cm}
\hbox{\hspace{0.215cm}\includegraphics[width=7.29truecm,height=3.5truecm]{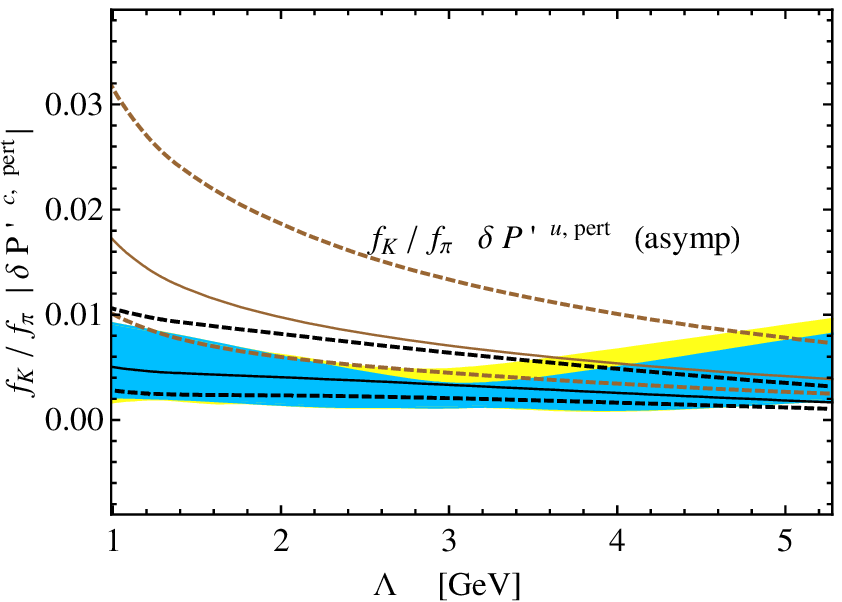}}\vspace{-.565cm}
\hbox{\hspace{0.262cm}\includegraphics[width=7.23truecm,height=3.5truecm]{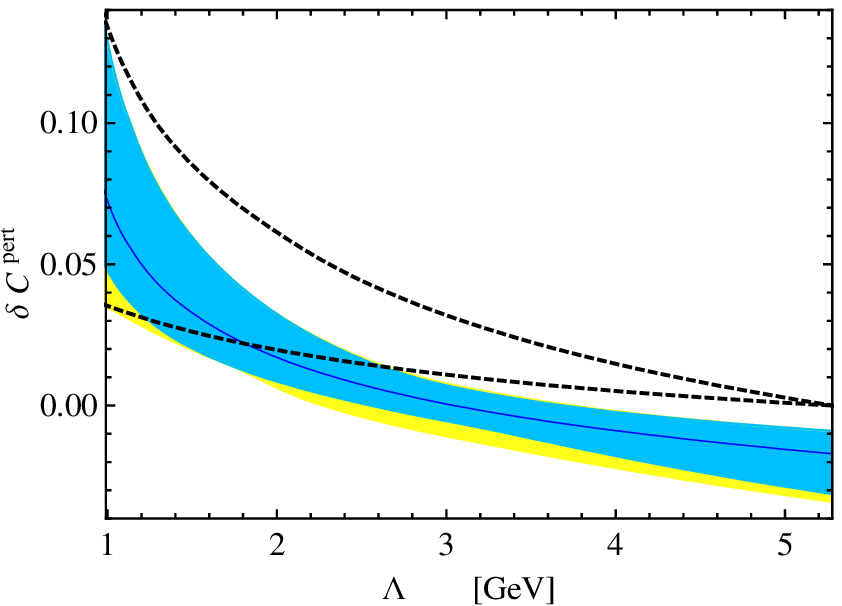}}}}
{\caption[1]{\label{fig:KPipert}
$B\to K\pi,\pi\pi$ perturbative PC's vs. $\Lambda$: solid curves for $\mu_b \!=\! m_b $, $\mu_h \! =\! \Lambda$, central values for other inputs; inner blue bands for $\mu_h = \Lambda$, $\mu_b  \in [m_b /2\,,m_B]$, inputs varied within errors; outer yellow bands include $\mu_h \in [ {\rm Max}[1\,{\rm GeV},\Lambda/2],\,m_B]$; dashed curves outline `outer bands' for asymptotic LCDA's, with black (brown) curves in (c) for $f_K /f_\pi \,\delta P^{\prime\,c\,(u)}$. Errors added
in quadrature.}} 
\end{figure}

In the non-perturbative limit  
($\Lambda \to \Lambda_{\rm QCD}$) the diagrams in Figs.~2a,b,e,, like those for $\delta F^{\rm pert.}$, 
give way to production of both light mesons from the soft overlap of an energetic approximately back-to-back pair of collinear partons.\footnote{Perturbative quark loops in the non-perturbative contributions will 
continue to cancel the leading $\log\mu_b $ scale dependence of the QCD penguin operator Wilson coefficients.}
This is indicated by quadratic dependences on $\log m_B /\Lambda $ and $\log \sqrt{s}/\Lambda$.
Our hypothesis is that this mechanism dominates in both the timelike form factor PC's and QCD penguin PC's.
Therefore, we would expect a similarly large soft-to-perturbative PC hierarchy in both,
given that the perturbative, as well as total PC's satisfy the power counting rules.
Indeed, comparison of
Figs.~4a,b,c with the fit results in Figs.~3a,b yields
\begin{equation}
|\delta P^{(\prime)\, c,\,{\rm n.p.}} /\delta P^{(\prime)\,c,\,{\rm pert.}}|\ge O(10)\,.
\label{eq:penguinsoftoverlap}
\end{equation}
This is consistent with the large hierarchy in Eq.~\ref{eq:nonpertvspert} for the timelike form factors, 
thus strengthening the case for a power correction explanation of the $B\to K\pi,\pi\pi$ puzzles.
Given the soft dominance of the penguin PC's, the origin of
$  \delta P^{\prime\,c} \,f_K /f_\pi  > \delta P^c$ and $\delta F_\pi  > \delta F_K$ would be the same: a larger 
$\pi\pi$ soft-overlap.  
There could be additional charm loop PC's 
in which the `loops' themselves are non-perturbative 
(corresponding to $c$ quarks near threshold), 
as has been suggested for LP penguins \cite{Bauer:2004tj}.
Our results indicate that neither are required by the data.

The hard spectator interaction diagrams in Figs.~2c,d,f contribute to the perturbative penguin power corrections (at levels well below those discussed above) and to the color-suppressed tree amplitudes $\delta C^{(\prime)\,\rm pert.} $ ($Q_1$ in Fig.~2c). In the non-perturbative limit ($\Lambda \!\to\! \Lambda_{\rm QCD}$) only the spectator quark light meson is produced via a soft-overlap,
as indicated by a linear dependence on $\log m_B/\Lambda$.  
Therefore, we might expect $|\delta C^{(\prime)\,,\,\rm n.p.} \!/\delta C^{(\prime)\,,\,\rm pert.}|$ to be smaller than
the penguin ratios in Eq.~(\ref{eq:penguinsoftoverlap}). This is consistent with Figs.~3c and 4d, albeit within large errors.
The requisite strong phase difference between $C$ and $T$ would have to be due to the soft-overlap.
The existence of strong phases from soft-overlaps would be confirmed by the measurement of a relative strong phase between different polarization amplitudes in 
$e^+e^- \to VV$ at the $\Upsilon (4S)$.

To summarize, we have argued that power corrections in the timelike form factors and QCD penguin amplitudes for $PP$ final states are dominated by soft-overlaps between pairs of energetic and approximately
back-to-back collinear partons.  This is supported by the large $\ge O(10)$ soft-to-perturbative power correction
hierarchies obtained in both cases.  The similarity
implies
that the magnitudes of the power corrections returned by fits to the $B \to K\pi,\pi\pi$ data ($1.4\sigma$ away from the Standard Model)
are natural, thus strengthening the case for a power correction explanation of the $B\to K\pi,\pi\pi$ puzzles.
\vspace{0.2cm}\newline
\noindent{\bf Note added:} After completion of this work, first reported in arXiv:0812.3162, we were informed by L.~Silvestrini that his group had also carried out a $K\pi$ PC fit, with similar predictions for $S_{K_s \pi^0}$, $C_{K_s \pi^0}$,  which appeared in \cite{Ciuchini:2008eh}. 
\vspace{0.2cm}\newline 
{\bf Acknowledgements:} We would like to thank T.~Becher, M.~Neubert, R.~Shrock, A.~Soni, and I.~Stewart for helpful discussions.  
This work was supported by DOE grant FG02-84-ER40153.


\begin{thebibliography}{99}

\bibitem{BBNS1}
%
M.~Beneke {\it et al.},
Nucl.\ Phys.\ B {\bf 606}, 245 (2001)
 \bibitem{CLEOc}
  T.~K.~Pedlar {\it et al.}  [CLEO Collaboration],
  Phys.\ Rev.\ Lett.\  {\bf 95}, 261803 (2005)
\bibitem{brodskylepage}   G.~P.~Lepage and S.~J.~Brodsky,
  Phys.\ Rev.\  D {\bf 22}, 2157 (1980)
   

\bibitem{ballLCDA}  
  P.~Ball, V.~M.~Braun and A.~Lenz,
  JHEP {\bf 0605}, 004 (2006)




\bibitem{milana}  
J.~Milana, S.~Nussinov, M.~G.~Olsson,  Phys.\ Rev.\ Lett.\  {\bf 71}, 2533 (1993)

\bibitem{gronaurosner1} 
M.~Gronau and J.~L.~Rosner,
  Phys.\ Rev.\  D {\bf 65}, 013004 (2002) [Erratum-ibid.\  D {\bf 65}, 079901 (2002)]
  
\bibitem{KPiHFAG} E.~Barberio {\it et al.}  [Heavy Flavor Averaging Group],
  http://www.slac.stanford.edu/xorg/hfag
\bibitem{benekejager1} 
  M.~Beneke and S.~Jager,
  Nucl.\ Phys.\  B {\bf 751}, 160 (2006)
 \bibitem{benekejager2} 
  M.~Beneke and S.~Jager,
  Nucl.\ Phys.\  B {\bf 768}, 51 (2007)


   
\bibitem{Gronau:1998fn}
  M.~Gronau, D.~Pirjol and T.~M.~Yan,
  Phys.\ Rev.\  D {\bf 60}, 034021 (1999)
  [Erratum-ibid.\  D {\bf 69}, 119901 (2004)]

\bibitem{inprep} M.~Duraisamy and A.L.~Kagan, in preparation


\bibitem{chiang}  C.~W.~Chiang and Y.~F.~Zhou,
  JHEP {\bf 0612}, 027 (2006)

\bibitem{fleischerzupan}   
R.~Fleischer {\it et al.},
  arXiv:0806.2900 [hep-ph]
\bibitem{gronaurosner2}   
M.~Gronau and J.~L.~Rosner,
  Phys.\ Lett.\  B {\bf 666}, 467 (2008)
  
\bibitem{CKMFitter} CKMfitter Group (J. Charles et al.), 
Eur. Phys. J. C41, 1-131 (2005),
[hep-ph/0406184], updated results available at http://ckmfitter.in2p3.fr

%
\bibitem{Bauer:2004tj}
  C.~W.~Bauer, D.~Pirjol, I.~Z.~Rothstein and I.~W.~Stewart,
  Phys.\ Rev.\  D {\bf 70}, 054015 (2004)

\bibitem{Ciuchini:2008eh}
  M.~Ciuchini, E.~Franco, G.~Martinelli, M.~Pierini and L.~Silvestrini,
  Phys.\ Lett.\  B {\bf 674}, 197 (2009)
\end{thebibliography}
\end{document}